\documentclass{aa}
\usepackage{natbib}
\newcount\longrefs
\def\adaa{\ifnum\longrefs=1 {Advances in Astronomy and Astrophysics}\else 
                           {Adv. Astron.\ Astrophys.}\fi}
\def\aap{\ifnum\longrefs=1 {Astron.\ Astrophys.}\else 
                           {A\hbox{\rm \&}A}\fi}
\def\aapr{\ifnum\longrefs=1 {Astron.\ Astrophys.\ Rev.}\else 
                            {A\hbox{\rm \&}AR}\fi}
\def\aaps{\ifnum\longrefs=1 {Astron.\ Astrophys.\ Suppl.}\else 
                            {A\hbox{\rm \&}AS}\fi}
\def\aj{\ifnum\longrefs=1 {Astron.\ J.}\else 
                          {AJ}\fi} 
\def\ao{\ifnum\longrefs=1 {Applied Optics}\else 
                           {Appl.\ Opt.}\fi} 
\def\aspcs{\ifnum\longrefs=1 {Astron.\ Soc.\ Pacific Conf. Series}\else 
                           {ASP Conf.\ Ser.}\fi} 
\def\apj{\ifnum\longrefs=1 {Astrophys.\ J.}\else 
                           {ApJ}\fi} 
\def\apjl{\ifnum\longrefs=1 {Astrophys.\ J. Lett.}\else 
                            {ApJ}\fi} 
\def\aplett{\ifnum\longrefs=1 {Astrophys.\ J. Lett.}\else 
                            {ApJ}\fi} 
\def\apjs{\ifnum\longrefs=1 {Astrophys.\ J. Suppl.}\else 
                            {ApJS}\fi}
\def\apss{\ifnum\longrefs=1 {Astrophys.\ and Space Science}\else 
                            {Ap\&SS}\fi}
\def\araa{\ifnum\longrefs=1 {Ann.\ Rev.\ Astron.\ Astrophys.}\else 
                            {ARA\hbox{\rm \&}A}\fi}
\def\azh{\ifnum\longrefs=1 {Astronomicheskii Zhurnal}\else 
                            {Astron.\ Zhur.}\fi}
\def\baas{\ifnum\longrefs=1 {Bull.\ Am.\ Astron.\ Soc.}\else 
                            {BAAS}\fi}
\def\bain{\ifnum\longrefs=1 {Bull.\ Astronom.\ Institutes Netherlands}\else
                            {Bull.\ Astr.\ Inst.\ Neth.}\fi}
\def\gca{\ifnum\longrefs=1 {Geochim.\ Cosmochim.\ Acta}\else 
                           {Geochim.\ Cosmochim.\ Acta}\fi}
\def\geo{\ifnum\longrefs=1 {Geophysical Journal}\else 
                           {Geophys.\ J.}\fi}
\def\grl{\ifnum\longrefs=1 {Geophys.\ Res.\ Lett.}\else 
                           {Geoph.\ Res.\ Lett.}\fi}
\def\iaucirc{\ifnum\longrefs=1 {IAU Circulars}\else 
                          {IAU Circ.}\fi}
\def\ip{\ifnum\longrefs=1 {in press}\else 
                          {in press}\fi}
\def\jgr{\ifnum\longrefs=1 {J.\ Geophys.\ Res.}\else 
                           {J.\ Geophys.\ Res.}\fi}  
\def\jrasc{\ifnum\longrefs=1 {J.\ Royal Astron.\ Soc.\ Canada}\else 
                           {JRAS Can.}\fi}  
\def\mnras{\ifnum\longrefs=1 {Mon.\ Not.\ Roy.\ Astron.\ Soc.}\else 
                             {MNRAS}\fi} 
\def\nat{\ifnum\longrefs=1 {Nature}\else 
                           {Nat}\fi}
\def\pasj{\ifnum\longrefs=1 {Pub.\ Astron.\ Soc.\ Japan}\else 
                            {PASJ}\fi} 
\def\pasp{\ifnum\longrefs=1 {Pub.\ Astron.\ Soc.\ Pacific}\else 
                            {PASP}\fi} 
\def\physscr{\ifnum\longrefs=1 {Physica Scripta}\else 
                            {Phys.\ Scrip.}\fi} 
\def\planss{\ifnum\longrefs=1 {Planetary \& Space Science}\else 
                            {Plan. \& Space Sci.}\fi} 
\def\procspie{\ifnum\longrefs=1 {Proc.\ SPIE}\else 
                            {Proc.\ SPIE}\fi} 
\def\qjras{\ifnum\longrefs=1 {Quarterly J.\ Royal Astron.\ Soc.}\else 
                            {QJRAS}\fi} 
\def\sa{\ifnum\longrefs=1 {Soviet Astron..}\else 
                               {Sov.\ Astron.}\fi}
\def\skytel{\ifnum\longrefs=1 {Sky \& Telescope}\else 
                            {Sky \& Tel.}\fi} 
\def\solphys{\ifnum\longrefs=1 {Solar Phys.}\else 
                               {Solar Phys.}\fi}
\def\ssr{\ifnum\longrefs=1 {Space Science Rev.}\else 
                               {Space\ Sci.\ Rev.}\fi}

\def\bibfiles{/home/bwillems/latex/bibtex/bibliofile}   

\def\aareferences{\longrefs=0  \bibliographystyle{/home/bwillems/latex/bibtex/aabib}
             \bibliography{/home/bwillems/latex/bibtex/aajour,\bibfiles}}

\def\dutch{\def\refname{Referenties}\def\abstractname{Samenvatting}%
  \def\bibname{Bibliografie}\def\chaptername{Hoofdstuk}%
  \def\appendixname{Bijlage}\def\contentsname{Inhoudsopgave}%
  \def\listfigurename{Lijst van figuren}\def\listtablename{Lijst van tabellen}%
  \def\indexname{Index}\def\figurename{Figuur}\def\tablename{Tabel}%
  \def\partname{Deel}\def\enclname{Bijlage(n)}\def\ccname{Ter attentie van}%
  \def\headtoname{Aan}\def\headpagename{Pagina}%
  \def\today{\number\day\space\ifcase\month\or januari\or februari\or maart\or%
     april\or mei\or juni\or juli\or augustus\or september\or oktober\or%
     november\or december\fi \space\number\year}%
  \typeout{
              >>>>> use hlatex209 for Dutch hyphenation <<<<< 
         }}
\newcounter{onefig} \newcounter{fignumber}
\newcount\nocaptions \newcount\nofigures \newcount\figwidth
\newcount\viewgraphs
  \def\paper{}  \def\figlabel{} 
\long\def\nextfig#1{\setcounter{figure}{\value{fignumber}}
  \addtocounter{fignumber}{1}
  \ifnum \viewgraphs=1 \newpage \pagestyle{empty} \fi 
  \ifnum\value{onefig}=0 #1 \fi                 
  \ifnum\value{onefig}=\value{fignumber} #1 \fi}
\def\figwidths#1#2{\ifnum \nocaptions=1 #2mm \else #1mm \fi}  
\def\paper#1{}  
\long\def\plotfig#1#2{\ifnum \nofigures=1 \else #2 \fi}
\long\def\captiontext#1{\ifnum \nofigures=1 \raggedright \fi 
   \ifnum \nocaptions=1 \paper
     \ifnum \viewgraphs=0 
       \newline  \mbox{}\hrulefill\mbox{} \newline 
       \newline label:~\{\figlabel\} 
     \fi 
     \else \ifnum \nofigures=0 \fi 
   #1 \fi}

\newcount\panelwidth \newcount\panelheight 
\newcount\bxmin \newcount\bymin \newcount\bxmax \newcount\bymax
\newcount\tbxmin \newcount\tbymin
\newcount\tpanelwidth \newcount\tpanelheight \newcount\tpdif
\def\panelsize #1,#2;{\panelwidth=#1 \panelheight=#2}  
\def\setbb #1,#2;#3,#4;#5,#6;{
  \tbxmin=#1 \tbymin=#2    
  \bxmin=#3 \bymin=#4      
  \bxmax=#5 \bymax=#6}     
\def\barepanel #1{%
  \ifnum\panelheight=0 
    \tpdif=\bymax \advance\tpdif by -\bymin
    \multiply \tpdif by \panelwidth
    \tpanelheight=\tpdif
    \tpdif=\bxmax \advance\tpdif by -\bxmin
    \divide \tpanelheight by \tpdif
  \else \tpanelheight=\panelheight \fi
  \epsfig{file=#1,%
     bbllx=\bxmin bp,bblly=\bymin bp,bburx=\bxmax bp,bbury=\bymax bp,clip=,%
     width=\panelwidth mm,height=\tpanelheight mm}}
\def\labelypanel #1{
  \ifnum\panelheight=0 
    \tpdif=\bymax \advance\tpdif by -\bymin
    \multiply \tpdif by \panelwidth
    \tpanelheight=\tpdif
    \tpdif=\bxmax \advance\tpdif by -\bxmin
    \divide \tpanelheight by \tpdif
  \else \tpanelheight=\panelheight \fi
  \tpdif=\bxmax \advance\tpdif by -\tbxmin
  \tpanelwidth=\panelwidth \multiply \tpanelwidth by \tpdif
  \tpdif=\bxmax \advance\tpdif by -\bxmin
  \divide \tpanelwidth by \tpdif
  \epsfig{file=#1,%
    bbllx=\tbxmin bp,bblly=\bymin bp,bburx=\bxmax bp,bbury=\bymax bp,%
    clip=,width=\tpanelwidth mm,height=\tpanelheight mm}}
\def\labelxpanel #1{%
  \ifnum\panelheight=0 
    \tpdif=\bymax \advance\tpdif by -\bymin
    \multiply \tpdif by \panelwidth
    \tpanelheight=\tpdif
    \tpdif=\bxmax \advance\tpdif by -\bxmin
    \divide \tpanelheight by \tpdif
  \else \tpanelheight=\panelheight \fi
  \tpdif=\bymax \advance\tpdif by -\tbymin
  \multiply \tpanelheight by \tpdif
  \tpdif=\bymax \advance\tpdif by -\bymin
  \divide \tpanelheight by \tpdif
  \epsfig{file=#1,%
    bbllx=\bxmin bp,bblly=\tbymin bp,bburx=\bxmax bp,bbury=\bymax bp,%
    clip=,width=\panelwidth mm,height=\tpanelheight mm}}
\def\labelxypanel #1{%
  \ifnum\panelheight=0 
    \tpdif=\bymax \advance\tpdif by -\bymin
    \multiply \tpdif by \panelwidth
    \tpanelheight=\tpdif
    \tpdif=\bxmax \advance\tpdif by -\bxmin
    \divide \tpanelheight by \tpdif
  \else \tpanelheight=\panelheight \fi
  \tpdif=\bxmax \advance\tpdif by -\tbxmin
  \tpanelwidth=\panelwidth \multiply \tpanelwidth by \tpdif
  \tpdif=\bxmax \advance\tpdif by -\bxmin
  \divide \tpanelwidth by \tpdif 
  \tpdif=\bymax \advance\tpdif by -\tbymin 
  \multiply \tpanelheight by \tpdif
  \tpdif=\bymax \advance\tpdif by -\bymin
  \divide \tpanelheight by \tpdif
  \epsfig{file=#1,%
    bbllx=\tbxmin bp,bblly=\tbymin bp,bburx=\bxmax bp,bbury=\bymax bp,%
    clip=,width=\tpanelwidth mm,height=\tpanelheight mm}}

\def\CC{\par \vspace*{-2ex} \footnotesize \baselineskip=8pt \begin{verbatim}}

\long\def\startignore #1\stopignore{}   

\def\setlistparams{         
  \topsep=0.7ex                 
  \itemsep=0.7ex                
  \leftmargini=3ex}             
\setlistparams                  

\newcounter{alistindex}       

\newcounter{romenumnr}

\newlength{\minipagewidth}

\newsavebox{\boxcontent}
\newcommand{\ovalhead}[1]{
  \unitlength=1cm
  \sbox{\boxcontent}{\mbox{~~{#1}~~}}
  \begin{center}
    \ifdim\wd\boxcontent>6ex 
    \ifdim\wd\boxcontent<8cm 
    \begin{picture}(8,3) \thicklines     
      \put(4.0,0.8){\oval(8,1.6)} 
      \put(0.0,0.7){\parbox{8cm}{
         \begin{center} \usebox{\boxcontent} \end{center}}}
    \end{picture}
    \else \ifdim\wd\boxcontent<12cm 
    \begin{picture}(12,3) \thicklines     
        \put(6.0,0.8){\oval(12,1.6)} 
        \put(0.0,0.7){\parbox{12cm}{
           \begin{center} \usebox{\boxcontent} \end{center}}}
    \end{picture}
    \else
    \begin{picture}(16,3) \thicklines     
        \put(8.0,0.8){\oval(16,1.6)} 
        \put(0.0,0.7){\parbox{16cm}{
           \begin{center} \usebox{\boxcontent} \end{center}}}
    \end{picture}
    \fi \fi \fi
  \end{center}} 

\newcounter{headnr}            
\newcounter{subheadnr}[headnr]
\newcounter{subsubheadnr}[subheadnr]
\def\head #1\par{
  \stepcounter{headnr}                          
  \vspace{2ex} \noindent                        
  {\bf \theheadnr~~~~#1}\\[1ex] \noindent}      
\def\subhead #1\par{  
  \stepcounter{subheadnr}
  \vspace{1.3ex} \noindent
  {\bf \theheadnr.\arabic{subheadnr}~~~#1}\\[0.3ex] \noindent}
\def\subsubhead #1\par{
  \stepcounter{subsubheadnr}
  \vspace{1.0ex} \noindent
  {\bf \theheadnr.\arabic{subheadnr}.\arabic{subsubheadnr}~~~#1}\\ \noindent}

\font\dropfont= cmr12 scaled \magstep5
\def\dropcap#1#2{{\noindent
    \setbox0\hbox{\dropfont #1}\setbox1\hbox{#2}\setbox2\hbox{(}%
    \count0=\ht0\advance\count0 by\dp0\count1\baselineskip
    \advance\count0 by-\ht1\advance\count0by\ht2
    \dimen1=.5ex\advance\count0by\dimen1\divide\count0 by\count1
    \advance\count0 by1\dimen0\wd0
    \advance\dimen0 by.25em\dimen1=\ht0\advance\dimen1 by-\ht1
    \global\hangindent\dimen0\global\hangafter-\count0
    \hskip-\dimen0\setbox0\hbox to\dimen0{\raise-\dimen1\box0\hss}%
    \dp0=0in\ht0=0in\box0}#2}

\def\level #1 #2#3#4{$#1 \: ^{#2} \mbox{#3} ^{#4}$}   

\def\mathstacksym#1#2#3#4#5{\def#1{\mathrel{\hbox to 0pt{\lower 
    #5\hbox{#3}\hss} \raise #4\hbox{#2}}}}

\mathstacksym\lta{$<$}{$\sim$}{1.5pt}{3.5pt} 
\mathstacksym\gta{$>$}{$\sim$}{1.5pt}{3.5pt} 
\mathstacksym\lrarrow{$\leftarrow$}{$\rightarrow$}{2pt}{1pt} 
\mathstacksym\lessgreat{$>$}{$<$}{3pt}{3pt} 

\usepackage{graphics}
\bibpunct{(}{)}{,}{a}{}{,}

\begin{document}
\headnote{Research Note}
\title{The effect of stellar compressibility and non-resonant dynamic
  tides on the apsidal-motion rate in close binaries}
\author{B.\ Willems\inst{1} \and A. Claret\inst{2}}
\offprints{B.\ Willems}
\institute{Department of Physics and Astronomy, The Open University,
Walton Hall, Milton Keynes, MK7 6AA, UK \\
\email{B.Willems@open.ac.uk}
\and
Instituto de Astrof\'{\i}sica de Andaluc\'{\i}a, CSIC, Apartado 3004,
E-18080 Granada, Spain \\
\email{claret@iaa.es} }
\titlerunning{Apsidal motion due to non-resonant dynamic tides}
\date{Received date; accepted date}

\abstract{The paper is devoted to the analysis of the systematic
  deviations between the classical formula for the rate of secular
  apsidal motion in close binaries and the formula established within
  the framework of the theory of dynamic tides. The basic behaviour is
  found to be governed by the dependency of the forcing frequencies on
  the orbital period and the rotational angular velocity, irrespective
  of the stellar model. For binaries with short orbital periods 
  a good agreement between the classical apsidal-motion formula and
  the formula established within the framework of the theory of
  dynamic tides is only possible for certain values of the
  rotational angular velocity. 
}

\maketitle

\section{Introduction}

In close binary systems of stars, the perturbation of the external
gravitational field due to the tidal and rotational distortions of the
component stars yields a periodic shift of the position of the
periastron. The effect was first studied by \citet{Rus1928}, and
subsequently by \citet{Cow1938} and \citet{Ste1939}. The latter
author derived a formula for the rate of secular apsidal motion based
on the assumption that the orbital period is long in comparison to the
periods of the free oscillation modes of the component stars. The
validity of this assumption has been investigated extensively by,
e.g., \citet{PP1980}, \citet{Sme1991}, \citet{Qua1996},
\citet{SWV1998}, and \citet{SW2001}. \citet{SW2001} found the
classical apsidal-motion formula to be valid up to high orbital
eccentricities as long as the orbital and rotational angular
velocities remain small. For binaries with higher orbital and
rotational angular velocities, the apsidal-motion rates predicted by
the classical formula deviate from those predicted by the theory of
dynamic tides due the effects of the compressibility of the stellar
fluid and the possibility of resonances between dynamic tides and free
oscillation modes of the component stars.

One of the most uncertain parameters in the analysis of the dynamics
of a close binary is the rotational angular velocity of the
component stars. \citet{CW2002} therefore considered the effects of
dynamic tides on the apsidal-motion rate as a function of the
rotational angular velocity for a carefully selected sample of
eclipsing binaries with accurate determinations of the stellar masses
and radii and the apsidal-motion rate. The authors found that even for
binaries with short orbital periods, a good agreement between the
classical apsidal-motion formula and the formula established within
the framework of the theory of dynamic tides is possible for certain
values of the rotational angular velocity.

In this research note, we aim to shed more light on the results found by
\citet{CW2002} by examining more closely the role of the forcing
frequencies and the effects of the compressibility of the stellar
fluid on the rate of secular apsidal motion due to the tidal distortions
of close binary components. In Sects.~2 and~3, we recall the
basic assumptions and the equations governing the rate of secular
change of the longitude of the periastron in the classical framework
and in the framework of the theory of dynamic tides. In Sect.~4, we
illustrate the behaviour of the relative deviations between the
classical apsidal-motion formula and the formula established within
the framework of the theory of dynamic tides as a function of the
rotational angular velocity in the case of an evolved
$5\,M_\odot$ main-sequence star. In Sects.~5 and~6, we unravel this
behaviour by looking at the effects of the forcing frequencies and the
stellar model separately. In the final section, we briefly summarise
our conclusions.

\section{Basic assumptions}

Consider a close binary system of stars with masses $M_1$ and $M_2$
that are orbiting around each other under the influence of their
mutual gravitational force. Let $P_{\rm orb}$ be the orbital period,
$a$ the semi-major axis, and $e$ the orbital eccentricity. We assume
the first star to rotate uniformly around an axis perpendicular to the
orbital plane with an angular velocity $\vec{\Omega}$ whose magnitude
is small in comparison to the stellar break-up speed $\vec{\Omega}_c$.
The companion star is treated as a point mass.

Furthermore, we denote by $\vec{r} = \left( r,\theta,\phi \right)$
a system of spherical coordinates with respect to an orthogonal frame
of reference that is corotating with the star. The potential governing
the tidal force exerted by the companion can then be expanded in terms
of unnormalised spherical harmonics $Y_\ell^m(\theta,\phi)$ and in
Fourier series in terms of multiples of the companion's mean motion
$n$ as
\begin{eqnarray}
\lefteqn{\varepsilon_T\, W \left( \vec{r},t \right) = -
  \varepsilon_T\, {{G\, M_1} \over R_1}\,
  \sum_{\ell=2}^4 \sum_{m=-\ell}^\ell \sum_{k=-\infty}^\infty}
  \nonumber \\
& & c_{\ell,m,k}\, \left( {r \over R_1} \right)^\ell
  Y_\ell^m (\theta,\phi)\, \exp \left[ {\rm i}
  \left( \sigma_T\, t - k\, n\, \tau \right) \right]
  \label{pot}
\end{eqnarray}
\citep[e.g.][]{Pol1990,SWV1998}. In this expansion, $G$ is the
Newtonian constant of gravitation, $R_1$ is the equilibrium radius of
the uniformly rotating star, $\varepsilon_T=\left(R_1/a\right)^3
M_2/M_1$ is a small dimensionless parameter, $\tau$ is a time of
periastron passage, and $\sigma_T = k\, n + m\, \Omega$ is a forcing
angular frequency with respect to the corotating frame of
reference. For a definition of the Fourier coefficients
$c_{\ell,m,k}$, we refer to \citet{SWV1998}. They are equal to
zero for odd values of $\ell+|m|$ and, for a given eccentricity,
decrease rapidly with increasing values of $k$.
 
For the remainder of the paper, we restrict ourselves to the
contributions of the dominant terms associated with $\ell=2$ in the
expansion of the tide-generating potential.

\section{The apsidal-motion rate due to the tidal deformations of
binary components}

The most commonly used formula for the rate of secular change of the
longitude of the periastron due to the tidal deformations of the
components of a close binary was derived by \citet{Cow1938} and
\citet{Ste1939} on the assumption that the orbital period is long in
comparison with the free harmonic periods of the component
stars. The rate of secular apsidal motion is then given by
\begin{equation}
\left( {{d \varpi} \over {dt}} \right)_{\rm classical}
  = \left({R_1\over a}\right)^5\, {M_2\over M_1}\,
  {{2\, \pi} \over P_{\rm orb}}\,
  k_2\, 15\, f\left({e^2}\right),  \label{strn:1}
\end{equation}
where $\varpi$ is the longitude of the periastron, $k_2$ is the
classical apsidal-motion constant, and
\begin{equation}
f\left(e^2\right) = \left({1-e^2}\right)^{-5}\,
  \left({1 + {3\over 2}\, e^2 + {1\over 8}\, e^4 }\right).
  \label{strn:2}
\end{equation}

A formula for the rate of secular change of the longitude of the
periastron taking into account the effects of dynamic tides was
derived by \citet{SWV1998} by adding the contributions to the
apsidal-motion rate stemming from the various terms in Expansion
(\ref{pot}) of the tide-generating potential. The resulting formula
takes the form
\begin{eqnarray}
\lefteqn{\left( {{d \varpi} \over {dt}} \right)_{\rm dyn} =
  \left({R_1 \over a}\right)^5 {M_2 \over M_1}\,
  {{2\, \pi} \over P_{\rm orb}}\, \bigg[
  2\, k_2\, G_{2,0,0} + 4\, \sum_{k=1}^\infty }
  \nonumber \\
 & & \left( F_{2,0,k}\, G_{2,0,k} + F_{2,2,k}\,
  G_{2,2,k} + F_{2,-2,k}\, G_{2,-2,k} \right) \bigg],
  \label{aps:1}
\end{eqnarray}
where the constants $F_{2,m,k}$ render the response of the star to the
various forcing angular frequencies $\sigma_T$, and the coefficients
$G_{2,m,k}$ are functions of the orbital eccentricity. For a
definition of the constants $F_{2,m,k}$ and the coefficients
$G_{2,m,k}$, we refer to \citet{SWV1998} and \citet{SW2001}. The
coefficients $G_{2,m,k}$ decrease rapidly with increasing values of
$k$.

For binaries with short orbital periods or rapidly rotating component
stars, the rates of secular apsidal motion predicted by the classical
formula deviate from the corresponding rates predicted by the formula
established in the framework of the theory of dynamic tides due to the
increased role of the stellar compressibility at higher forcing
frequencies and due to resonances of dynamic tides with free
oscillation modes of the component stars \citep{SW2001}. The
extent of the deviations depends on the evolutionary stage of the star
due to the increase of the stellar radius and the redistribution of
the mass as the star evolves on the main sequence \citep{WC2002}.  In
the present investigation, we focus on the role of the forcing
frequencies and the compressibility of the stellar fluid. In analogy
to previous investigations \citep{SW2001,WC2002,CW2002}, we study the
deviations between the apsidal-motion rates predicted by the classical
formula and the rates predicted by the formula taking into account the
effects of dynamic tides by means of the relative difference
\begin{equation}
\Delta = {{\displaystyle \left( {d \varpi/dt} \right)_{\rm classical}
  - \left( {d\varpi/dt} \right)_{\rm dyn}} \over {\displaystyle
  \left( {d\varpi/dt} \right)_{\rm  dyn}}}.  \label{Delta}
\end{equation}

\section{Relative deviations for a $5\,M_\odot$ main-sequence star}

We illustrate the behaviour of the relative differences $\Delta$ as a
function of the rotational angular velocity $\Omega$ in the case
of a $5\,M_\odot$ main-sequence star with a central hydrogen
abundance $X_c=0.36$ and a radius $R_1=3.8\,R_\odot$. Since the
rotational angular velocity $\Omega$ is assumed to be small, we
neglect the effects of the Coriolis force and the centrifugal
force for the determination of the constants $F_{2,m,k}$. The tides
then correspond to those of a non-rotating spherically symmetric
equilibrium star \citep[for more details see, e.g.,][]{WC2002}.

The resulting relative deviations $\Delta$ are presented in
Fig.~\ref{deltaM5} for the orbital eccentricity $e=0.4$ and for
orbital periods ranging from 5.0 to 12.5 days. For convenience, the
rotational angular velocity $\Omega$ is expressed in units of the
orbital angular velocity $\Omega_P$ at the periastron of the relative
orbit. The peaks in the curves correspond to resonances between
dynamic tides and free oscillation modes. Since we are here
interested in the systematic trend caused by the effects of the
forcing frequencies and the compressibility of the stellar fluid
rather than in a detailed treatment of resonant dynamic tides, we did
not attempt to find all possible resonances for each orbital 
configuration. For a more detailed treatment of the resonances,
we refer the reader to \citet{SW2001} and \citet{WC2002}.

\begin{figure}
\resizebox{\hsize}{!}{\includegraphics{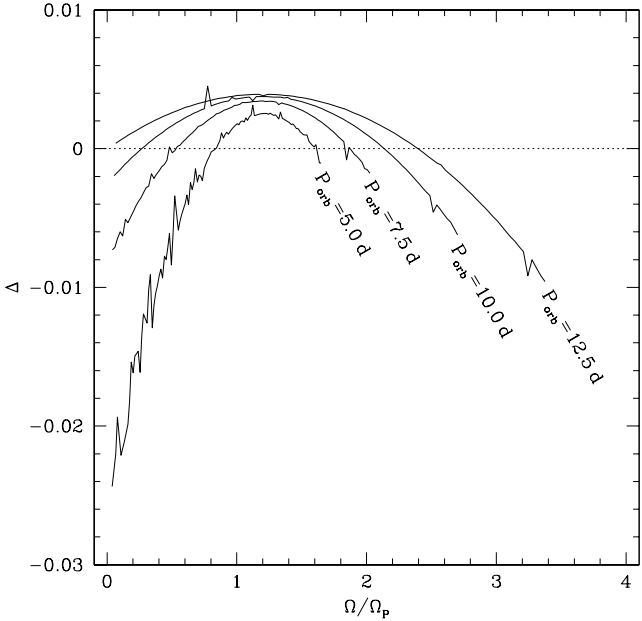}}
\caption{ The relative differences $\Delta$ for the evolved
  $5\,M_\odot$ main-sequence model, the orbital eccentricity $e=0.4$,
  and the orbital periods $P_{\rm orb}=5.0,7.5,10.0,12.5$ days.}
\label{deltaM5}
\end{figure}

In the limiting case where $\Omega \rightarrow 0$, the forcing
frequencies $\sigma_T$ are determined solely by the mean motion
$n$. As shown by \citet{SW2001}, the relative differences $\Delta$ are
then generally negative and larger in absolute value for shorter
orbital periods. For larger values of $\Omega$, the relative
differences $\Delta$ increase, reach a positive maximum, and
subsequently decrease to become negative.  Hence, we find two values
of the rotational angular velocity for which the relative differences
$\Delta$ become zero. These values depend on both the period and the
eccentricity of the orbit, and do not necessarily correspond to
synchronised rotation with the orbital angular velocity at the
periastron of the relative orbit.  We will further scrutinise this
behaviour in the following section.

\section{The role of the forcing frequencies}

Due to the expansion of the tide-generating potential in Fourier
series in terms of multiples of the companion's mean motion, the
tidally distorted star is subjected to an infinite number of forcing
angular frequencies $\sigma_T$ [see Eq.~(\ref{pot})]. For a given
orbital eccentricity, the largest differences between the classical
apsidal-motion formula and the formula established within the
framework of the theory of dynamic tides can be expected to occur when
high forcing frequencies are associated with large Fourier
coefficients $c_{2,m,k}$. Vice versa, the smallest differences can be
expected to occur when the dominant terms in the expansion of the
tide-generating potential all have small forcing frequencies. 

A similar reasoning applies to Expansion~(\ref{aps:1}) which renders
the apsidal-motion rate within the framework of the theory of dynamic
tides. Since the system of differential equations governing
non-resonant dynamic tides in components of close binaries depends on
the square of the forcing frequency, the deviations of the constants
$F_{2,m,k}$ from the classical apsidal-motion constant $k_2$ are of
the order of $\sigma_T^2$. Because of this, \citet{SW2001} and
\citet{CW2002} were able to approximate the systematic deviations
between the classical apsidal-motion formula and the formula
established within the framework of the theory of dynamic tides by
second-degree polynomials in the companion's mean motion $n$ or the
star's rotational angular velocity $\Omega$. In the present context,
this property implies that the differences between the two formulae
can be expected to be large when high forcing frequencies are
associated with large coefficients $G_{2,m,k}$.

In order to shed some more light on this, we examine the behaviour of
the function
\begin{equation}
S_1 \left( P_{\rm orb}, \Omega \right) = \sum_{m=-2}^2
   \sum_{k=0}^\infty \sigma_T^2\,\, G_{2,m,k}^2.
   \label{S1}
\end{equation}
The function mimics the behaviour described above in the sense that it
will be large when high forcing frequencies are associated with large
coefficients $G_{2,m,k}$.

The variations of $S_1$ as a function of the rotational angular
velocity $\Omega$ are displayed in Fig.~\ref{ffo1} in the case of the
orbital eccentricity $e=0.4$ and orbital periods ranging from 5.0 to
12.5 days. For given values of $e$ and $P_{\rm orb}$, the function
$S_1$ initially decreases with increasing values of
$\Omega$. The decrease is caused by the decrease of the forcing
frequencies associated with $m=-2$, which generally provide the
dominant contributions to $S_1$. As $\Omega$ becomes larger, $S_1$
reaches a minimum when the forcing frequencies associated with the
largest coefficients $G_{2,-2,k}$ are close to zero. The frequencies
associated with $m=-2$ subsequently become more and more negative,
resulting in an indefinite increase of the function $S_1$. 

\begin{figure}
\resizebox{\hsize}{!}{\includegraphics{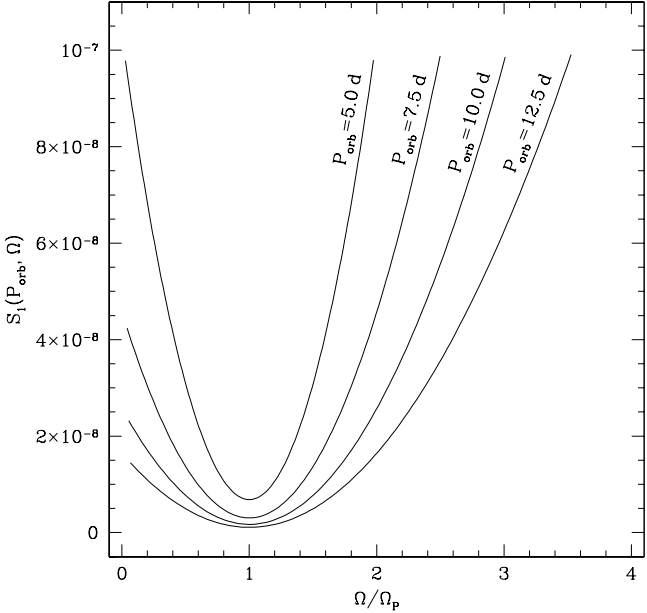}}
\caption{The variation of $S_1 \left( P_{\rm orb}, \Omega
  \right)$ as a function of the rotational angular velocity $\Omega$
  for the same orbital eccentricity and orbital periods as
  considered in Fig.~\ref{deltaM5}.}
\label{ffo1}
\end{figure}

As a preliminary conclusion we can thus say that the opposite of the
function $S_1$ shows the same qualitative behaviour as the 
relative differences $\Delta$ shown in Fig.~\ref{deltaM5}. In the next
section, we will make this even clearer by adding some information on
the stellar model to the definition of $S_1$.

\section{The role of the stellar model}

The response of a particular stellar model to the tidal forcing of a
close companion is determined by the constants $F_{2,m,k}$.
For low-frequency, non-resonant dynamic tides in stars with a radiative
envelope, these constants can be approximated by their
asymptotic representation
\begin{equation}
F_{2,m,k}^{\rm (a)} = k_2 - {R_1^3 \over {G\, M_1}}\,
  {{\sigma_T^2} \over {12}} \left[ \left({{d\,\xi_2}\over
  {dr}}\right)_{R_1}\!\! + 2\, {{\xi_2\left({R_1}\right)}\over R_1}
  \right], \label{F2mka}
\end{equation}
where $\xi_2(r)$ is a solution of the homogeneous
second-order differential equation
\begin{equation}
{{d^2 \xi_2} \over {dr^2}}
  + 2 \left[ {1 \over {m(r)}} {{d\,m(r)} \over {dr}}
  - {1 \over r} \right] {{d\,\xi_2} \over {dr}}
  - {4 \over r^2}\, \xi_2 = 0
  \label{stat:1}
\end{equation}
\citep{SWV1998}. In the latter equation, $m(r)$ is the mass contained
inside the sphere with radius $r$. The solution of Eq.~(\ref{stat:1})
must furthermore remain finite at $r=0$ and satisfy the 
boundary condition
\begin{equation}
\left({{d\,\xi_2} \over {dr}}\right)_{R_1}\!\!
  + {{\xi_2\left({R_1}\right)}\over R_1}  = 5.
  \label{stat:2}
\end{equation}

The asymptotic approximation for the constants $F_{2,m,k}$ is valid
for forcing frequencies $\sigma_T \ll \sqrt{ GM_1/R_1^3}$. Higher
forcing frequencies, such as those associated with large values of
$k$, require additional terms in the asymptotic
approximation. However, the second-order approximation given by
Eq.~(\ref{F2mka}) is sufficient to further clarify the qualitative
behaviour of the relative differences $\Delta$ observed in  
Fig.~\ref{deltaM5}. 

We proceed in a similar way as for the introduction of the function
$S_1$ in the previous section, but now take into account the structure
of the stellar model by defining a function $S_2$ as
\begin{equation}
S_2 \left( P_{\rm orb}, \Omega \right) = \sum_{m=-2}^2
  \sum_{k=0}^\infty F_{2,m,k}^{\rm (a)} G_{2,m,k}^2.
  \label{S2a}
\end{equation}
The function can be rewritten in terms of $S_1$ as
\begin{eqnarray}
\lefteqn{S_2 \left( P_{\rm orb}, \Omega \right) = k_2\! \sum_{m=-2}^2
  \sum_{k=0}^\infty G_{2,m,k}^2 }  \nonumber \\
 & & - {1 \over {12}}\, {R_1^3 \over {G\, M_1}}\,
  \left[ \left({{d\xi_2} \over {dr}}\right)_{R_1}\!\!
  + 2\, {{\xi_2\left({R_1}\right)}\over R_1} \right]
  S_1 \left( P_{\rm orb}, \Omega \right). \label{S2b}
\end{eqnarray}
For a given stellar model and a given orbital eccentricity, the first
term is then seen to be constant, while the second term depends
on the orbital period and the rotational angular velocity through the
function $S_1$.

The variations of $S_2$ as a function of the rotational angular
velocity $\Omega$ are displayed in Fig.~\ref{ffo2} for the same
$5\,M_\odot$ main-sequence model used in Sect.~4.  The qualitative
behaviour of $S_2$ is essentially the same as that of $-S_1$. In
addition, the inclusion of the model dependent factors introduces two
zeros in the curves, similar to the behaviour of the relative
differences $\Delta$ displayed in Fig.~\ref{deltaM5}.

\begin{figure}
\resizebox{\hsize}{!}{\includegraphics{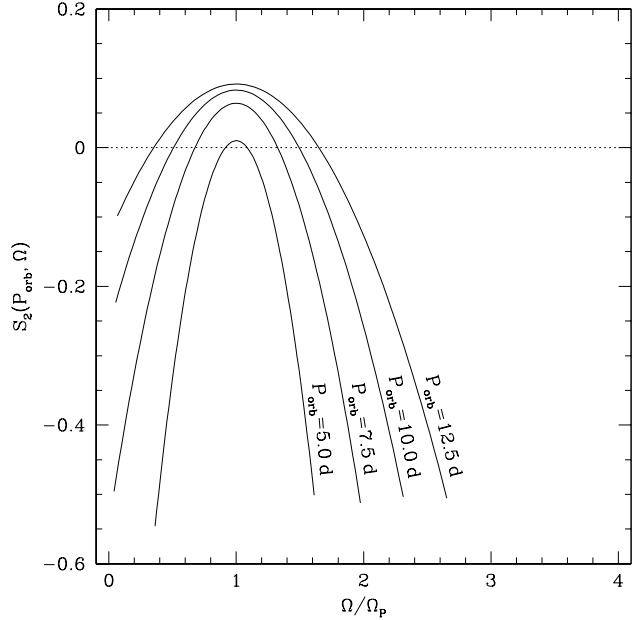}}
\caption{The variation of $S_2 \left( P_{\rm orb}, \Omega
  \right)$ for the evolved $5\,M_\odot$ main-sequence model and
  the same orbital eccentricity and orbital periods as
  considered in Fig.~\ref{deltaM5}.}
\label{ffo2}
\end{figure}

\section{Summary}

In this research note, we scrutinised the behaviour of the systematic
relative deviations between the classical formula for the rate of
secular apsidal motion and the formula established within the
framework of the theory of dynamic tides as a function of the often
uncertain rotational angular velocity of the component stars.
We found that when the effects of resonant dynamic tides are
neglected, the basic behaviour of the relative differences is governed
by the dependency of the forcing frequencies on the orbital period and
the rotational angular velocity, irrespective of the stellar
model. For a given value of the orbital eccentricity, a good agreement
between the classical apsidal-motion formula and the formula
established within the framework of the theory of dynamic tides is
possible even for binaries with short orbital periods, but only
for certain values of the rotational angular velocity.

\begin{acknowledgements}
BW and AC acknowledge the support received by the British Particle
Physics and Astronomy Research Council (PPA/G/S/1999/00127) and the
Spanish DGI (AYA2000-2559), respectively. 
\end{acknowledgements}

\aareferences

\end{document}